\documentclass{sigchi}

\usepackage[T1]{fontenc}

\usepackage{balance}
\usepackage{amsmath}
\usepackage{graphicx}
\usepackage{booktabs}
\usepackage{colortbl}
\usepackage[usenames,dvipsnames]{xcolor}
\usepackage{bm}
\usepackage{booktabs}
\usepackage{textcomp}
\usepackage{microtype} 
\usepackage{ccicons} 
\usepackage{times}

\PassOptionsToPackage{pdfpagelabels=false}{hyperref}

\usepackage{todonotes}

\usepackage{ucs}
\usepackage[utf8x]{inputenc}

\usepackage[]{color}
\usepackage[breaklinks]{hyperref}

\hypersetup{colorlinks=true, linkcolor=Black, citecolor=Black, filecolor=Blue,
    urlcolor=Blue, unicode=true}

\usepackage[all]{hypcap}  

\usepackage{url}
\makeatletter
\g@addto@macro{\UrlBreaks}{\UrlOrds}
\makeatother

\begin{document}

\conferenceinfo{CHI}{'16 San Jose, California, USA}

\title{Surviving an ``Eternal September''\\\vspace{0.2em}How an Online Community Managed a Surge of Newcomers}

\numberofauthors{3}
\author{%
  \alignauthor{Charles Kiene\\
    \affaddr{University of Washington}\\
    \affaddr{Seattle, Washington}\\
    \email{ckiene@uw.edu}}
  \alignauthor{Andrés Monroy-Hernández\\
    \affaddr{Microsoft Research}\\
    \affaddr{Redmond, Washington}\\
    \email{amh@microsoft.com}}
  \alignauthor{Benjamin Mako Hill\\
    \affaddr{University of Washington}\\
    \affaddr{Seattle, Washington}\\
    \email{makohill@uw.edu}}\\
}

\toappear{Permission to make digital or hard copies of part or all of this work
for personal or classroom use is granted without fee provided that copies are
not made or distributed for profit or commercial advantage and that copies bear
this notice and the full citation on the first page. Copyrights for third-party
components of this work must be honored. For all other uses, contact the
Owner/Author.\\ Copyright is held by the owner/author(s).\\
CHI'16, May 07-12, 2016, San Jose, CA, USA\\
ACM 978-1-4503-3362-7/16/05.\\
\href{http://dx.doi.org/10.1145/2858036.2858356}{http://dx.doi.org/10.1145/2858036.2858356}}

\maketitle

\begin{abstract}
 We present a qualitative analysis of interviews with participants in the NoSleep community within Reddit where millions of fans and writers of horror fiction congregate. We explore how the community handled a massive, sudden, and sustained increase in new members. Although existing theory and stories like Usenet's infamous ``Eternal September'' suggest  that large influxes of newcomers can hurt online communities, our interviews suggest that NoSleep survived without major incident. We propose that three features of NoSleep allowed it to manage the rapid influx of newcomers gracefully: (1) an active and well-coordinated group of administrators, (2) a shared sense of community which facilitated community moderation, and (3) technological systems that mitigated norm violations. We also point to several important trade-offs and limitations.
\end{abstract}



\section{Introduction}

\begin{figure}
\centering
\includegraphics[width=\columnwidth]{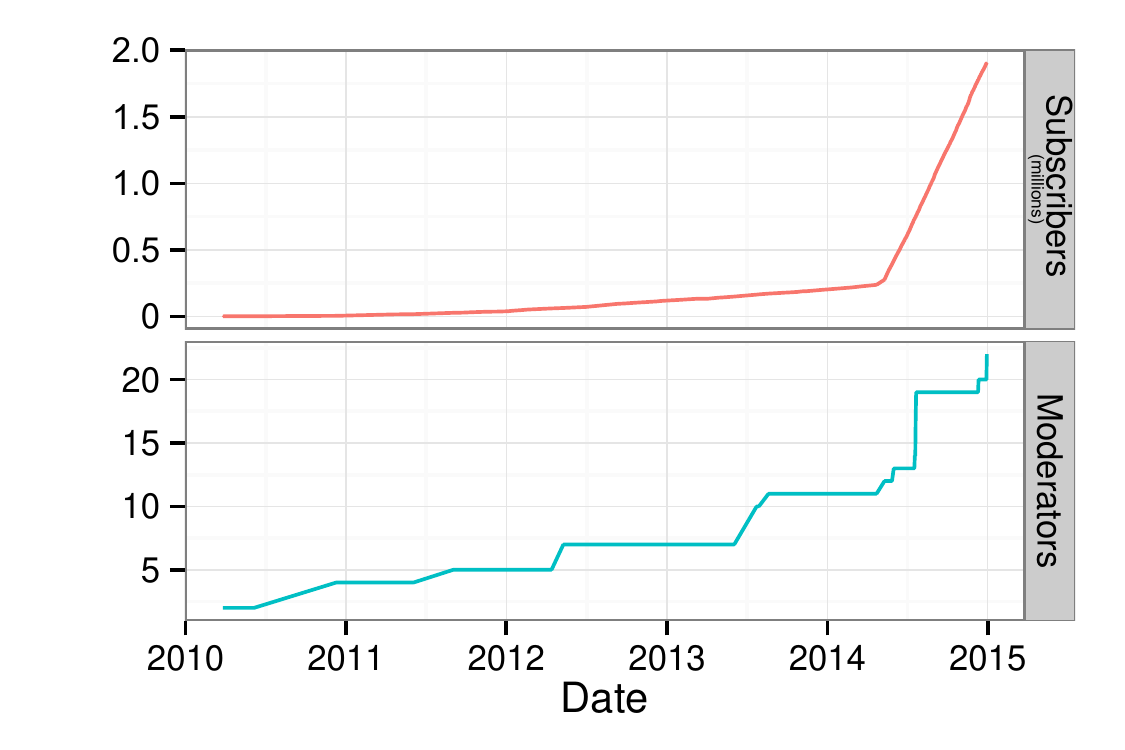}
\caption{Plots of numbers of subscribers (in millions), and moderators over the life of the community. Data retrieved from snapshots available at archive.org}
\label{fig:growth}
\end{figure}

NoSleep\footnote{\href{https://www.reddit.com/r/nosleep}{https://www.reddit.com/r/nosleep}} is an online community hosted on the social media website Reddit where people share, rate, and comment on original horror stories in an immersive environment. From its inception in May 2010 to May 2014, NoSleep grew organically to more than 240,000 subscribers. On May 7, 2014, Reddit's administrators added NoSleep to a list of communities that every new Reddit user is automatically subscribed to. Less than a month later, NoSleep's subscriber-base had doubled and it has continued to grow at this pace (see Figure \ref{fig:growth}). Although theory suggests that large influxes of newcomers will challenge, disrupt, and can even destroy online communities, NoSleep appeared to manage this growth without major negative effects.
Using a grounded theory-based analysis of interviews of NoSleep members, writers, and moderators, we suggest that NoSleep was able to survive and thrive through this massive influx of newcomers because it had created systems that ensured a high degree of adherence to the community's norms and that minimized the effect of violations.
Our findings also point to several important trade-offs and limitations.

\section{Background}

The problem of how to attract newcomers is one of the most important challenges for builders of online communities \cite{kraut_building_2012}. A growing body of work considers how leadership, framing, competition, and membership overlap can play an important role in attracting newcomers \cite{schweik_internet_2012,  zhu_selecting_2014, kraut_building_2012, zhu_impact_2014}. However, relatively little research has considered the challenges faced by communities that successfully manage to attract large numbers of new members.

The problems caused by successfully attracting large numbers of newcomers are often invoked in terms of an ``Eternal September''
\cite{grossman_net.wars_1998}. In the 1980s and early 1990s, Usenet participants frequently complained about inexperienced university students joining their communities in September, the beginning of the North American academic year. When America Online connected to Usenet in 1994 and unleashed a large and unremitting stream of new users, Usenet denizens complained of a ``September'' that never ended and irrevocably damaged their community.

Social computing research suggests that influxes of newcomers can disrupt communities in many ways including increased activity leading to information overload \cite{butler_membership_2001, jones_information_2004} and alienation among established users \cite{jeffries_systers:_2005}. Most prominently, research has suggested that newcomers cause disruptions because they do not know and do not follow community norms \cite{kraut_building_2012}. As a result, new users can disrupt conversations, contribute low quality content, and annoy existing users. For example, Gorbatai has shown that newcomers tend to make low quality edits to Wikipedia which require fixes by established editors \cite{gorbatai_aligning_2012}.

Because most social computing research treats community growth as a desirable goal, most studies of newcomer activity have focused on how newcomers can be deterred by sanctions leveled at good faith norm violations. For example, studies have shown that Wikipedia editors whose work is undone are less likely to contribute in the future \cite{halfaker_rise_2013, halfaker_dont_2011, piskorski_testing_2013}. In response, designers have created welcoming spaces within existing communities \cite{morgan_tea_2013} and tools to identify and welcome good faith contributors \cite{halfaker_snuggle:_2014}.

This work has limited relevance for communities in NoSleep's position in May 2014 whose challenge was widespread norm violation by a surge of newbies. Norm violations included linking to external content in ways that are normative elsewhere in Reddit, marking work as fiction or non-fiction, and misusing the Reddit voting system. Perhaps most importantly, many new users violated a very strong norm of suspended disbelief that requires all commenters to act as if stories are factual. For example, asking if something is true, or even complimenting the author for a ``nice story,'' is not permitted on NoSleep. Violation of this norm by newcomers is disruptive to NoSleep's immersive environment and has historically been treated as a serious threat to the community.

Although the social computing literature points to a deep toll taken by this type of widespread norm violation \cite{kraut_building_2012}, NoSleep seems to have largely survived its own Eternal September. The NoSleep participants we interviewed suggested that, ``\emph{it's gotten bigger but not necessarily worse}'' (P1) and that, ``\emph{if you went on today, you would see all the comment threads just filled with people going along with it and just enjoying the experience.}'' (P5). Our analysis asks: How did NoSleep survive, and even thrive, through its Eternal September?

\section{Methodology}

Because large influxes of newcomers are largely unstudied in the social computing literature, the phenomenon is well-suited to qualitative theory-building. Consequently, we adopted a grounded theory interview-based approach \cite{charmaz_constructing_2006}. We recruited members of NoSleep in two ways. First, we posted several messages in NoSleep-related forums on Reddit describing the study and requesting that participants contact a member of our research team with information about their individual role and experience in NoSleep. We identified a set of roles (e.g., moderator, writer) and other dimensions (e.g., gender, ex-user) as theoretically important. As is common in grounded theory, we used these dimensions to build a sample that was stratified but did not attempt to be statistically representative \cite{trost_statistically_1986}. In some cases, we reached out to individuals who we felt would have illuminating perspectives (e.g., a founding member of the community). In total, we interviewed 12 subjects as described in Table \ref{tab:subjects}. All participants were compensated with a \$10 Amazon gift card.

\begin{table}[]
\centering
\begin{tabular}{cllll}
\hline
ID & Gender & Role & Joined & Length\\
\hline
1 & Female & Lurker & 2013 & 46 min \\
2 & Female & Active & 2011 & 36 min \\ 
3 & Female & Moderator & 2010 & 41 min \\
4 & Female & Active & 2012  & 42 min \\ 
5 & Male & Founder / Moderator & 2010 & 52 min \\
6 & Female & Lurker & 2010 & 46 min \\
7 & Male & Ex-Active / Writer & 2013 & 77 min \\
8 & Female & Lurker & 2012 & 41 min \\
9 & Male & Moderator / Writer & 2012 & 62 min \\
10 & Female & Lurker & 2013 & 24 min \\
11 & Female & Active & 2013 & 48 min \\
12 & Male & Moderator & 2010 & 43 min \\
\hline 
\end{tabular}
\caption{List of study participants with participant ID as used in this paper, gender, primary role(s) in NoSleep, year that they joined the community, and the length of our interview. The term ``active'' indicates an intense combination of reading, voting, and commenting. ``Lurker'' indicates reading and voting and was associated with less deep involvement in the community.}
\label{tab:subjects}
\end{table}

Subjects were interviewed over the phone, or via audio/video chat for an average of 47 minutes. Although interviews were semi-structured and open-ended, our protocol was designed to elicit feedback on the large influx of users in May 2014 and is provided in the supplemental materials. All interviews were recorded and transcribed. 
Following the methodology laid out by Charmaz \cite{charmaz_constructing_2006}, the first author coded each interview using a series of both inductive codes emergent from the text and deductive codes identified by theory. Transcripts and codes were discussed by all the authors and transcripts were recoded in an iterative fashion. Ultimately, codes were grouped into themes that were elaborated in iteratively created memos.

\section{Findings}

Our analysis of coded interviews suggests that NoSleep survived and thrived during its Eternal September because it was equipped with three interconnected systems that ensured a high degree of adherence to NoSleep's norms while minimizing the effect of  violations. We present these findings in terms of three propositions and suggest that these attributes could help other online communities survive and thrive in the face of large influxes of newcomers. We also describe ways in which the individuals we interviewed reflected on the trade-offs introduced by these features.
 
\subsection{Proposition 1. Consistent Enforcement by Leaders}

Participants attributed NoSleep's success in the face of a large influx of new users to the exceptional responsiveness and effectiveness of the community's moderators (``mods'') who wield broad authority to remove content and ban users.
Figure \ref{fig:growth} shows that there were a dozen moderators in May 2014 and that the size of the group has accelerated since then. P1 commented on the quality of moderation saying, \emph{``the NoSleep mods really do a lot as far as keeping everyone not just on track and following rules, but also like keeping everyone interested and active.''} NoSleep's moderators were described as a group of community insiders who were committed to, and effective in, keeping the community stable and sustainable. 

Moderator work primarily involves rule enforcement and sanctioning. Our subjects commented on the consistency and strictness of NoSleep's moderators and described these qualities as an important component of NoSleep's success in the face of massive growth. Several NoSleep moderators active in other Reddit subcommunities explained that NoSleep had both the strictest and the most consistently enforced rules they had encountered. For example, moderator P3 described how she would enforce community norms at the expense of suppressing friendly conversation:
\emph{``If people come on and they say `that [story] was really great,' that's technically breaking the rules. But, you have to be a real jerk -- that's why you're like `nope you can't -- you cannot praise good writing.'\thinspace''}

Although moderators like P3 mentioned examples of the social and emotional challenges of enforcing NoSleep's norms, they also felt that their work was essential to maintain the 
stability and immersiveness of the community. Although frequently described as inflexible, subjects also described community leadership as engaged, fair, and legitimate. Echoing the experience of several subjects, P10 described at length how a moderator's interventions helped her learn how to effectively navigate and interact within the community.

In comparison to other communities, the NoSleep moderators were described as extremely organized and engaged. Interviews with moderators revealed a large and sophisticated behind-the-scenes infrastructure including an entire private Reddit community used by moderators to communicate, collaborate, and coordinate with each other to ensure that their actions were responsive and consistent. P3 described the usefulness of this private subreddit: ``\emph{We put out drafts of moderator announcements and everybody suggests additions or things that should be changed, so that when we go out and moderate the community -- we're really able to show a unified front.}''

\subsection{Proposition 2. Moderation By Community Members}

The community members we interviewed suggested that widespread engagement in norm enforcement by ``normal'' community members was critically important to handling newcomers. They also suggested that this type of community regulation was only made possible by a strong shared sense of community. For example, there was a striking degree of shared understanding across all participants on what constituted good NoSleep stories. Although subjects could reflect on their individual taste in stories, many respondents echoed P6's claims that excellent NoSleep stories should include ``\emph{a strong character voice}'' and be ``\emph{something that's almost believable}.'' Although many newcomers adopt more off-the-cuff styles, nearly every subject interviewed mentioned style and grammar as criteria that they use to judge the quality of stories on NoSleep. Some of this knowledge is made explicit in documentation on the site.

The degree to which this knowledge reflects a shared sense of community was also visible in the way community members described working together to address examples of users violating NoSleep's norms. Described as ``burying'' material, subjects described collectively rating norm-violating content and comments as low quality (i.e., ``downvoting'') so that it becomes hidden by Reddit's interface. P1 explained how she approached comments left by newcomers unaware of the suspension of disbelief norm, ``\emph{that's when you're like `All right, if we all downvote this, it will go away. It will be like it never happened}.''' Although P1 did not write content or comments herself, she expressed a feeling of empowerment to act with the community to preserve its norms that was nearly universal among our interviewees.

Although rules are rigid, P8, P9 and P10 each reflected on the way that the community's sense of purpose and shared goals made it possible to identify attempts to game its system, and on the community's ability to ``correct themselves'' in these cases. Because the number of moderators with explicit authority is limited, this sense of community made collective action among ``normal'' members possible, scaled effectively, and was able to both minimize damage from, and educate, a sustained influx of newcomers. Because users could work to ``bury'' norm-violating material, there was much less pressure on moderators to act quickly and in every case of a violated norm.

\subsection{Proposition 3. Technological Systems Maintaining Norms}
               
Participants suggested that the technological affordances of NoSleep were a third important factor in the community's smooth growth.
Several technologies mentioned included basic functionality of the Reddit platform that facilitates community moderation. This included Reddit's voting system giving all members the ability to quickly vote content up or down.
As P1 explained, ``\emph{in order to be an active member of the community, you just need to hit a button}.''
A related tool proved by Reddit facilitates ``peer reporting'' of problematic content which is presented to moderators who can then hide content and contact users.
Moderator P12 explained that, ``\emph{if you have individuals committed enough to ... not break immersion, just by clicking a report button on comments that do [break immersion], it brings it into the moderator queue so that we can make them disappear}.''

Building on this system, NoSleep employs a tool called ``AutoModerator'' that automatically detects rule violations and sanctions violators \cite{morris_reddit_2012}.
Although not visible to many users, several moderators we interviewed explained that the tool, also provided by Reddit, finds and flags problematic content and communicates with moderators in ways that obviates the need for action on many straightforward moderation issues.

Interviewees also credited Reddit's functionality that allows newcomers to edit and improve stories or comments after discovering they have inadvertently broken a rule or deviated from the community's norms and then to resubmit their content. They also pointed to a feature that issues reminders of norms at points of action including an HTML placeholder attribute in the comment box below stories that reminds users of rules before commenting. One moderator described a tool used by the community called ``post throttling'' which limits users to one story submission every 24 hours. P3 explained that throttling was used to reduce the threat of a newcomer disrupting the community and to provide a limit on the effect of a trend of newbies posting ``series'' of stories to garner visibility and popularity. 

Of course, these technologies rely on social infrastructure to be effective in ways that highlight the interrelated nature of our propositions. Community voting and peer reporting rely on an engaged set of community members with a shared sense of the community as well as an active set of moderators who can effectively remove flagged or downvoted content and sanction repeat violators. Our interviewees suggested that NoSleep effectively combined these three features into a socio-technical system that was able to maintain community standards through a sustained influx of newcomers.

\subsection{The Cost of Strong Regulation}
All three propositions point to systems facilitating strong norm enforcement. Although described as important for managing the influx of new users, these systems were not described  as universally positive or costless.
For example, a rule requiring stories to be believable was elaborated in the aftermath of the influx of new users to explicitly bar supernatural stories (e.g., stories that involve demons or vampires). Although subjects acknowledged that this pushed newcomers toward creating believable stories, it also annoyed some established users who felt they could navigate the fine line between supernatural content and believability. For example, P3  explained that, ``\emph{the rules had been made tighter because of the new subscribers, and that sometimes doesn't allow them as much freedom}.'' Similarly, P7 felt that rules were, ``\emph{corralling younger users into acting or reading in a particular way...instead of doing what they want}.'' Frustrated by this experience, P7 explained that he no longer contributes to the community as frequently.

For others, tough rule enforcement was seen as having a negative effect on commenting and discussion by making the environment feel constrained and contrived. P11 described the difference between her experiences in NoSleep and other Reddit subcommunities, commenting on the rules that forbid any kind of criticism of stories:
\emph{``Most subreddits have comment sections that are kind of stream of consciousness—people tend to share their own experiences. On NoSleep, you don't see people really sharing their own experiences; you see people commenting specifically on the story…Again, because of the rules, you don't really see trolling. Kind of nice, kind of not nice ... I think the comment section is kind of -- I don't think it's very organic.''}
Through a strong system of rules and a complex socio-technical infrastructure to ensure that they are enforced, NoSleep was able to survive and thrive despite the weight of millions of newcomers.
However, the cost of NoSleep's survival was described by some interviewees as an uncomfortably strict environment that limited creativity. The systems described as facilitating NoSleep's rapid growth were also portrayed as providing strict limits on what it could grow to become.

\section{Discussion}

In one sense, our three propositions seem to be at odds with other social computing research. For example, a body of Wikipedia research has connected stronger systems of norms with inefficient bureaucracies that may cause communities growth to slow or even stop \cite{butler_dont_2008, suh_singularity_2009, jullien_rise_2015}. For example, Halfaker et al.~\cite{halfaker_rise_2013} connect increases in social and technical systems for norm-enforcement to lower rates of newcomer retention. In another sense, our propositions should not be unfamiliar to social computing researchers. For example, all three propositions can be found in some form in among Kraut and Resnick's \cite{kraut_building_2012} ``design claims'' and our contribution lies not in the discovery of these features but in our suggestion that they play a critical role in helping groups survive and thrive through what is often traumatic or catastrophic growth. We believe that techniques that minimize the effect of norm violations by newcomers may both help prevent communities from descending into chaos \emph{and} deter newcomers. Techniques that prevent short-term disaster may be inappropriate -- and difficult to change -- when growth slows.

Of course, our work is limited in many ways. One unavoidable limitation of our inductive  grounded theory approach is that findings may reflect the idiosyncrasies and biases of interviewees. For example, we were only able to recruit one user who described themselves as a former NoSleep member. As a result, our findings may reflect ``survivor bias'' where individuals less negatively affected by an event are the only people available to be interviewed. We gain some confidence in our findings by the fact that our participants did not describe any major exoduses of authors or moderators or major changes in the nature of the community. That said, we only present these findings as propositions for testing in future work.

Our findings offer several important implications for design. The first points toward the importance of emphasizing decentralized moderation. Although previous research has found this to be ``underprovisioned'' on Reddit as a whole \cite{gilbert_widespread_2013}, in NoSleep it seems to be sufficient.
A second implication is the importance of ensuring enough leadership capacity is available when an influx of newcomers is anticipated. Designers may benefit by focusing on tools to let existing leaders bring others on board and help them clearly communicate norms.
Finally, designers should support an ecosystem of accessible and appropriate moderator tools. During a widely reported Reddit uprising, a moderator of a different subcommunity complained that, ``the moderation tools we are given are severely lacking''  \cite{warzel_reddit_2015}.

Our interviews and analysis point to the importance of strong systems of norm enforcement made possible by leadership, community engagement, and technology. Although we propose that NoSleep's socio-technical infrastructure can provide a template for other communities facing similar challenges, we also suggest that they are not without trade-offs. Although not without qualification, NoSleep's example provides a model for how an Eternal September need not mean an inevitable march toward winter.

\bibliographystyle{SIGCHI-Reference-Format}
\bibliography{refs}

\end{document}